 \documentclass[iop]{emulateapj}

\usepackage{graphicx}
\usepackage{hyperref}

\slugcomment{Accepted to the Astronomical Journal}

\shorttitle{SN site IFU spectroscopy -- SN type-II progenitors}
\shortauthors{Kuncarayakti et al.}

\begin{document}


\title{Integral field spectroscopy of supernova explosion sites: constraining mass and metallicity of the progenitors -- II. Type II-P and II-L supernovae}


\author{Hanindyo Kuncarayakti}
\affil{Kavli Institute for the Physics and Mathematics of the Universe (WPI), Todai Institutes for Advanced Study, the University of Tokyo, 5-1-5 Kashiwanoha, Kashiwa, Chiba 277-8583, Japan}
\affil{Institute of Astronomy, Graduate School of Science, the University of Tokyo, 2-21-1 Osawa, Mitaka, Tokyo 181-0015, Japan}
\affil{Department of Astronomy, Graduate School of Science, the University of Tokyo, 7-3-1 Hongo, Bunkyo-ku, Tokyo 113-0033, Japan}
\email{hanindyo.kuncarayakti@ipmu.jp}

\author{Mamoru Doi}
\affil{Institute of Astronomy, Graduate School of Science, the University of Tokyo, 2-21-1 Osawa, Mitaka, Tokyo 181-0015, Japan}
\affil{Research Center for the Early Universe, the University of Tokyo, 7-3-1 Hongo, Bunkyo-ku, Tokyo 113-0033, Japan}

\author{Greg Aldering}
\affil{Physics Division, Lawrence Berkeley National Laboratory, 1 Cyclotron Road, Berkeley, CA 94720}

\author{Nobuo Arimoto}
\affil{National Astronomical Observatory of Japan, 2-21-1 Osawa, Mitaka, Tokyo 181-0015, Japan}
\affil{Subaru Telescope, National Astronomical Observatory of Japan, 650 North A\rq{ohoku} Place, Hilo, HI 96720}

\author{Keiichi Maeda}
\affil{Kavli Institute for the Physics and Mathematics of the Universe (WPI), Todai Institutes for Advanced Study, the University of Tokyo, 5-1-5 Kashiwanoha, Kashiwa, Chiba 277-8583, Japan}

\author{Tomoki Morokuma}
\affil{Institute of Astronomy, Graduate School of Science, the University of Tokyo, 2-21-1 Osawa, Mitaka, Tokyo 181-0015, Japan}

\author{Rui Pereira}
\affil{CNRS/IN2P3, Institut de Physique Nucl\'eaire de Lyon, 4 Rue Enrico Fermi, 69622 Villeurbanne Cedex, France}

\author{Tomonori Usuda}
\affil{Subaru Telescope, National Astronomical Observatory of Japan, 650 North A\rq{ohoku} Place, Hilo, HI 96720}

\author{Yasuhito Hashiba}
\affil{Institute of Astronomy, Graduate School of Science, the University of Tokyo, 2-21-1 Osawa, Mitaka, Tokyo 181-0015, Japan}
\affil{Department of Astronomy, Graduate School of Science, the University of Tokyo, 7-3-1 Hongo, Bunkyo-ku, Tokyo 113-0033, Japan}





\begin{abstract}
Thirteen explosion sites of type II-P and II-L supernovae in nearby galaxies have been observed using integral field spectroscopy, enabling both spatial and spectral study of the explosion sites. We used the properties of the parent stellar population  of the coeval supernova progenitor star to derive its metallicity and initial mass (c.f. Paper I). The spectrum of the parent stellar population yields the estimates of metallicity via strong-line method, and age via comparison with simple stellar population (SSP) models. These metallicity and age parameters are adopted for the progenitor star. Age, or lifetime of the star, was used to derive initial (ZAMS) mass of the star by comparing with stellar evolution models. With this technique, we were able to determine metallicity and initial mass of the SN progenitors in our sample. Our result indicates that some type-II supernova progenitors may have been stars with mass comparable to SN Ib/c progenitors. 
\end{abstract}


\keywords{supernovae: general -- stars: massive}



\section{Introduction}
Recent efforts in studying supernova progenitors have been greatly empowered by the availability of high-resolution archival images of the explosion site, enabling direct identification of the progenitor star before the explosion. However, among core-collapse supernovae (SNe), only the progenitors of type-II SNe have been directly detected and well characterized (see \citet{smartt09araa} for a review). On the other hand, theoretical predictions for SN progenitor stars have been proposed \citep[e.g.][]{heger03,eldridge04,georgy09} but those are not very well tested with observational data.

\citet{smartt09} presented a comprehensive study of nearby type II-P SNe with pre-explosion imaging. By comparing progenitor luminosity and colors in several passbands with theoretical stellar evolution models, the properties of the SN progenitor star were derived. Despite the fact that only a handful of SN II-P progenitors have good detections in several passbands, and the majority are only detected in few bands or even not detected, they were able to derive the mass range of SN II-P progenitors to be within $8.5^{+1}_{-1.5}$ and $16.5\pm1.5$~M$_\odot$.

SN II-L are thought to have lost more of their hydrogen envelope compared to the type-II-P. There are fewer successful detections of II-L progenitors (SN 2009kr \citep{eliasrosa10,fraser10}, SN 2009hd \citep{eliasrosa11}), but already these provide indications that SNe II-L are produced by stars more massive than II-P progenitors. However, the position of the SN II-L progenitors on the Hertzprung-Russell diagram implies that they were yellow supergiant stars shortly prior to the explosion. This introduces a problem since stellar evolution predictions do not expect a star to explode as a SN in this yellow supergiant stage. Considering the low number of progenitor detections it is necessary to study this issue further, either with the same or other methods. 

While pre-explosion progenitor detection offers the most direct method in constraining progenitor properties, it is greatly limited by the availability of usable pre-explosion data of the explosion site thus difficult to increase the statistics. Metallicity is usually assumed since it is not possible to be derived from imaging data alone. The use of proxies for metallicity determination is not uncommon. The determination of progenitor properties depends very much on the location of the purported star on the H-R diagram -- which is sensitive to uncertainties in luminosities and the final stages of stellar evolution. Furthermore, the detected progenitor could be confused with other nearby stars, a binary companion, or a small compact star cluster until post-explosion observation confirm that it has already disappeared after the explosion. Up to this point, other than the very nearby SN 1987A there are only few firm SN progenitor disappearances have been reported in the literature: SN IIn 2005gl \citep{galyam09}, SN IIb 1993J and SN II-P 2003gd \citep{maund09}, SN II-P 2008bk \citep{mattila10}. Alternative strategies in constraining progenitor properties from the local environment have frequently been implemented \citep[e.g.][]{leloudas11,anderson12,sanders12}, providing more insights into the nature of SN progenitors. The basic reasoning of this study is similar to that of \citet{gogarten09}, who estimated the initial progenitor mass of NGC 300 OT2008-1 to be 12--25 M$_\odot$ from the analysis of the stellar population within 50 pc from the transient.

In this paper and also the preceding paper in this series \citep[Paper I]{paper1} we report the results of our investigation of nearby SN explosion sites using integral field spectroscopy (IFS). Taking advantage of IFS, the SN explosion sites could be studied spatially and spectrally to reveal the nature of the stellar populations present there. The parent stellar population of the SN progenitor star provides metallicity and age estimates of the SN progenitor, assuming it was coeval with the parent cluster. 

Stars are born in clusters \citep[][also see \citet{bressert10} for a discussion on how the adopted cluster definition may change the fraction of stars born in clusters, between $\sim45$--90\%]{lada03}, therefore it is possible to derive age and metallicity of a star based on its parent star cluster. Furthermore, all massive stars may have been born within clustered environments \citep{pz10}. As velocity dispersion inside a star cluster is typically a few km/s \citep{bastian06,pz10}, or a few pc/Myr, the short-lived progenitor star is expected to be still not far from the parent cluster if it is unbound. In the two papers we use the terms star cluster, H\texttt{II} region, and OB association interchangeably to refer to a stellar population. The age, or lifetime, of the progenitor corresponds to the initial (ZAMS) mass of the star, since the evolution of a single star is mainly governed by the amount of mass it had at its birth time. With this method we are able to put constraints on the metallicity and initial mass of several core-collapse SN progenitors.

The paper is organized as follows. We present the observations and data analysis in Section \ref{data}, followed by the description and results for each explosion site in Section \ref{sites}. We discuss the estimate of contamination in Section \ref{contam} and the overall result is discussed in Section \ref{discu}. Finally, the paper is summarized in Section \ref{summ}.

\section{Data acquisition and analysis method}
\label{data}
The method of our data acquisition and analysis is the same as described in \citet[Paper I]{paper1}. Here we repeat the description of data acquisition and analysis. We used the Asiago Supernova Database \citep{barbon99} to select our samples. Broadband images of SN host galaxies with radial velocity 3000 km/s or less were inspected visually using ALADIN\footnote{\url{http://aladin.u-strasbg.fr/aladin.gml}} to find SNe associated with bright stellar populations. We used DSS and SDSS images for this purpose, and also additionally the published SN environment images by \citet{boffi99}. The study of \citet{boffi99} was initially meant to observe SN light echoes but detected star clusters instead.

In this visual inspection we selected SNe with close association with a bright knot at the explosion site. The knots are interpreted as the parent stellar populations of the SN progenitor stars. We do not expect to have any preference towards very young stellar populations, since the selection was based on broadband images rather than, for example, H$\alpha$ or $U$-band images which are dominated by light from stellar populations of very young age. With this selection method we prevent to include age bias in our sample of SN explosion sites. Table~\ref{tabobs} lists our SN site targets and observations. The fifth column of Table~\ref{tabobs} shows the positional uncertainty of each SN. The reasons for each estimates are given in the description of each explosion sites. Typically Hubble Space Telescope (HST) observation of a SN gives sub-arcsecond accuracy.

The SuperNova Integral Field Spectrograph (SNIFS; \citet{aldering02}, \citet{lantz04}) attached to University of Hawaii's 2.2 m telescope (UH88) atop Mauna Kea was used to observe the explosion sites. The observation was carried out in total of five nights, in 2010--2011. Observing condition was generally near-photometric with few intermittent clouds. With SNIFS we were able to observe a 6.4"$\times$6.4" field on the sky spatially and spectroscopically within the full range of optical wavelength between 3300 -- 9700 \AA. The spatial resolution of each spaxel (spatial pixel) is determined by the lenslet size, corresponding to 0.43 arcsec on the sky. 
A photometric channel in SNIFS is used to accurately place objects in the IFU. The positional accuracy with which a given (RA, Dec) coordinates is placed on the IFU is about 0.2 arcsec. The scale and rotation of the SNIFS IFU is well measured thus should not produce much additional astrometric uncertainty. We note that the telescope pointing did not always center on the SN positions but rather off-set to include the SN position and the nearby star clusters within the IFU field of views (see Figures~\ref{sp70G} onwards). The approximate SN positions indicated on the IFU fields in the Figures are estimated using Aladin from the SN position--cluster center offset and orientation on the broadband images.
SNIFS is controlled by remote operation and a fully-dedicated pipeline processes the raw data to produce final wavelength- and flux-calibrated ($x,y,\lambda$) datacubes. \citet{aldering06} present the outline of data reduction process which is similar to the description in \S 4 of \citet{bacon01}. 
During the observing run we obtained integral field spectroscopy of 16 nearby type II-P and II-L SN sites. However, in this paper only those with reliable SN positional uncertainties are presented (13 SN sites).

\begin{deluxetable*}{lcllclrlcc}
\tabletypesize{\scriptsize}
\tablecaption{Target SN sites IFU observations}
\tablewidth{0pt}
\tablehead{
\colhead{SN} & \colhead{Type} & \colhead{RA2000} & \colhead{Dec2000} & \colhead{$\sigma_{\alpha,\delta}$} &\colhead{Galaxy ($N_{\textrm{SN}}$)} &
\colhead{$d$/Mpc\tablenotemark{a}} & \colhead{Obs. date\tablenotemark{b}} & \colhead{Exposure} &
\colhead{Seeing} 
}
\startdata
1970G & II-L & 14:03:00.83 & +54:14:32.8 & $\pm0.2$" & NGC 5457 (3)   &   6.9 &  2011 Mar 10  & 1800 s $\times1$  &   1.3"  \\
2009hd & II-L &  11:20:16.99 & +12:58:46.3  & $\pm0.01$" & NGC 3627 (4)  & 10.0  & 2011 Mar 15 & 1800 s $\times2$ & 0.6" \\
2009kr & II-L & 05:12:03.30 & $-$15:41:52.2 & $\pm0.02$"  & NGC 1832 (2) & 26.2 & 2011 Mar 11 & 1800 s $\times2$ & 1.3" \\
1961I & II &  12:22:00.44 & +04:28:13.3 & $\pm2$" & NGC 4303 (6) & 16.4 & 2011 Mar 15 & 1800 s $\times2$ & 1.0" \\
1994L & II &  09:20:08.8 & $-$16:32:28 & $\pm1$"  & NGC 2848 (1) & 27.7 & 2011 Mar 11 & 1800 s $\times2$ & 1.3" \\
1999gi & II-P &   10:18:16.66 & +41:26:28.2 & $\pm0.02$"  & NGC 3184 (6) & 11.9 & 2011 Mar 11 & 1800 s $\times2$ & 1.0" \\
1999gn & II-P &  12:21:57.04 & +04:27:45.7 & $\pm0.1$"  & NGC 4303 (6) & 16.4 & 2011 Mar 15 & 1800 s $\times2$ & 0.8" \\
2002hh & II-P &   20:34:44.29 & +60:07:19.0 & $\pm0.1$" & NGC 6946 (9) & 5.9 & 2010 Aug 1 & 1800 s $\times1$ & 0.8" \\
2003ie & II-P &  12:03:18.15 & +44:31:34.6 & $\pm0.17$"  & NGC 4051 (3) & 14.5  & 2011 Mar 13 & 1800 s $\times2$ & 0.8" \\
2004am & II-P &   09:55:46.61 & +69:40:38.1 & $\pm0.1$"  & NGC 3034 (3) & 3.7 & 2011 Mar 10 & 1800 s $\times2$ & 1.1" \\
2004dj & II-P &   07:37:17.02 & +65:35:57.8 & $\pm0.1$"  & NGC 2403 (3) & 3.5 & 2011 Mar 10 & 1800 s $\times2$ & 1.3" \\
2005ay & II-P &   11:52:48.07 & +44:06:18.4 & $\pm0.1$" & NGC 3938 (3) & 17.4 & 2011 Mar 15 & 1800 s $\times2$ & 1.4" \\
2008bk & II-P &   23:57:50.42 & $-$32:33:21.5 & $\pm0.05$"  & NGC 7793 (1) & 4.1 & 2010 Aug 1 & 1800 s $\times1$ & 0.8" \\
\enddata
\tablenotetext{a}{Mean redshift-independent distance from NED (http://ned.ipac.caltech.edu)}
\tablenotetext{b}{Hawaiian Standard Time (UTC -- 10)}
\label{tabobs}
\end{deluxetable*}

The final datacubes were measured and analysed using IRAF\footnote{IRAF is distributed by the National Optical Astronomy Observatories, which are operated by the Association of Universities for Research in Astronomy, Inc., under cooperative agreement with the National Science Foundation.}. The datacubes are regarded as stacks of images taken at different wavelengths. For each wavelength 'slice', flux density was measured by performing aperture photometry on objects in the field using the task \textit{apphot} in IRAF. The seeing FWHM was used as the aperture radius, or in cases where the object is too close to one another or to the field edge, a smaller radius was used. Sky subtraction was done using annular apertures around the object; in most cases the annulus is larger than the field thus in effect only small part of it was used to measure the sky. The position of the object along wavelength was traced, thus eliminating the effect of differential atmospheric refraction \citep[DAR;][]{filippenko82}.

Arranging each photometry measurements in wavelength, the spectrum of each object was obtained. Subsequent spectral analysis was done using IRAF/\textit{splot}. The nebular emission lines and stellar absorption lines were measured by fitting a Gaussian curve. Prior to line measurement for metallicity determination, the stellar continuum was removed from the spectrum by fitting a polynomial function. In the equivalent width measurement a polynomial function was also used to fit and normalize the continuum. 

We use oxygen abundance as the measure of metallicity. This was done using O3N2 and N2 indices of \citet[][hereafter PP04]{pp04}, requiring the observation of the line ratios [N\texttt{II}]$\lambda$6584/H$\alpha$ (for N2 index determination),  and [O\texttt{III}]$\lambda$5007/H$\beta$ (for O3N2 index determination). We use the value of solar oxygen abundance of 12+log(O/H) = 8.66 \citep{asplund04}, following PP04. In the cases where only N2 determination is possible, the resulting metallicity is adopted; otherwise metallicity is averaged from O3N2 and N2 determinations. PP04 mentioned the $1\sigma$ error in metallicity from N2 determination to be $\pm$0.18 dex. When metallicity is determined by both N2 and O3N2, the quoted errors represent the upper and lower bounds of the metallicity in Z$_\odot$ unit.

\begin{figure}[Ht!]
\plotone{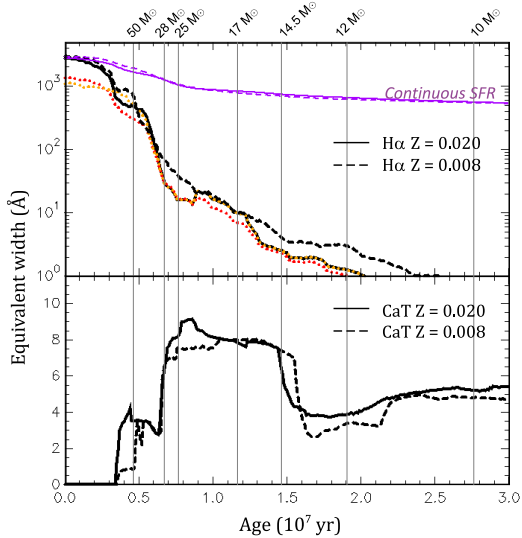}
\caption{(Upper panel:) time evolution of H$\alpha$ equivalent width, Starburst99 SSP. Single-burst model is represented with black line color and continuous star formation with purple; both use standard Salpeter IMF with $\alpha = 2.35$, M$_{\textrm{up}} = 100$ M$_\odot$. Dotted lines represent single-burst solar-metallicity models with different IMFs; red is for $\alpha = 3.30$, M$_{\textrm{up}} = 100$ M$_\odot$, orange is for $\alpha = 2.35$, M$_{\textrm{up}} = 30$M $_\odot$.
(Lower panel:) evolution of Ca\texttt{II} triplet equivalent width, single-burst Starburst99 SSP with IMF $\alpha = 2.35$, M$_{\textrm{up}} = 100$M$\odot$.
The lifetimes of single stars of different initial masses at solar metallicity according to Padova models are indicated with vertical grey lines.}
\label{hacat}
\end{figure}

The age of the stellar population was determined by comparing age indicators in the spectrum with simple stellar population (SSP) models from Starburst99 \citep{leitherer99}. We assume an instantaneous-burst population with standard Salpeter IMF ($\alpha=2.35$). For the age indicator, we primarily use the equivalent width (EW) of H$\alpha$ emission, and also of the near-infrared Ca\texttt{II} triplet (CaT) at $\lambda\lambda$8489, 8542, 8662 as the secondary indicator. Their evolution with SSP age is presented in Figure \ref{hacat}. While H$\alpha$ show an almost monotonic behaviour with age, CaT is quite degenerate and only EW values around 6 \AA$ $ or larger is useful for age determination. As age solution for that range is not single-valued, again H$\alpha$ is needed to constrain the solution. For example, at CaT EW = 6 \AA, the possible age solutions are 7 or 14.5 Myr. The presence of H$\alpha$EW will be used to decide whether the solution is 7 or 14.5 Myr. Despite this degeneracy, CaT is very useful since it is a good indicator for the presence of red supergiant stars.

The error for equivalent width measurements was estimated from the signal-to-noise ratio of the continuum part of the spectrum. The age of the stellar population is equal to the age of the SN progenitor, due to the assumption of coevality. This corresponds to the lifetime of the progenitor star, which is determined by its initial mass. Padova stellar evolution models of \citet{bressan93} for solar metallicity (Z = 0.02) and \citet{fagotto94} for 0.4 solar metallicity (Z = 0.008) were used to estimate the initial mass of the star from its lifetime, for each respective metallicities. The dividing line between using Z = 0.02 and Z = 0.008 model is the observed oxygen abundance of 0.7 (O/H)$_\odot$, corresponding to 12+log(O/H) = 8.50. In Paper I we demonstrated that the selection of SSP model for age determination does not affect the result of progenitor initial mass significantly, generally with consistency better than 20--30 \%.

\section{The explosion sites}
\label{sites}

\medskip
\subsection{SNe II-L sites}

\medskip
\begin{flushleft}
\textit{1. \objectname{SN 1970G} site} \\
\end{flushleft}
\citet{fesen93} reported the optical rediscovery of the SN almost 22 years after maximum light. The reported SN position was accurate to within 0.2". SN-circumstellar matter interaction was suggested as the energy source of the late-time emission. SN 1970G has also been detected in late times in other wavelengths: radio \citep{stockdale01}, and X-ray \citep{immler05}.

\begin{figure*}[ht!]
\plotone{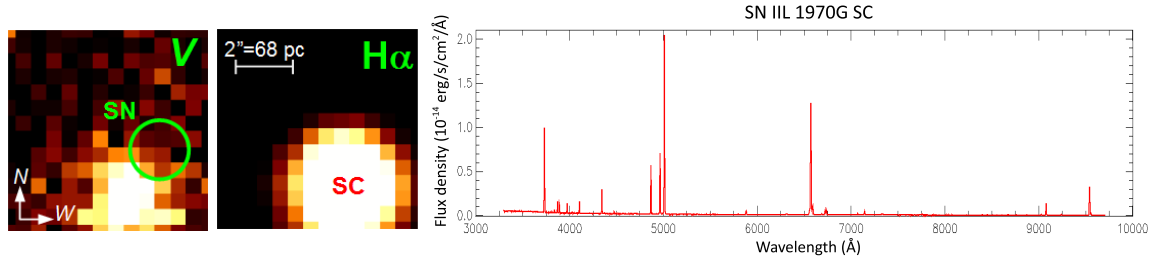}
\caption{Left panel: SN 1970G site reconstructed IFU FoV. SN position within 1 arcsec error radius is indicated by a circle. Approximate linear scale corresponding to 2 arcsec is also indicated; this scale is calculated from the host galaxy distance only thus does not take into account the projection effect and host galaxy inclination. "SC" indicates the host star cluster. Right panel: extracted spectrum of the star cluster.}
\label{sp70G}
\end{figure*}

Our observation shows that the spectrum of the cluster is very similar in appearance with the host of SN 1961U -- a blue continuum rising towards UV region and dominated by strong emission lines of ionized gas (Figure \ref{sp70G}). High-order Paschen emission lines are present along with helium lines. No Wolf-Rayet (WR) star signature could be found. We determined the metallicity of the site as around half solar. At this metallicity the age determined from H$\alpha$ is 3.4 Myr, which corresponds to the lifetime of a very massive star exceeding $\sim100$~M$_\odot$.

\medskip
\begin{flushleft}
\textit{2. \objectname{SN 2009hd} site} \\
\end{flushleft}
SN 2009hd exploded in a nearby spiral galaxy M66 (NGC 3627). The study by \citet{eliasrosa11} provides a comprehensive investigation of SN 2009hd evolution and the nature of its progenitor star. Using pre-explosion HST images, they detected a possible progenitor in F814W images but not in F555W filter. The purported progenitor might have been a luminous red or yellow supergiant with initial mass $\lesssim20$~M$_\odot$ based on their analysis, with positional uncertainty in the order of 0.01". 

\begin{figure*}[ht!]
\plotone{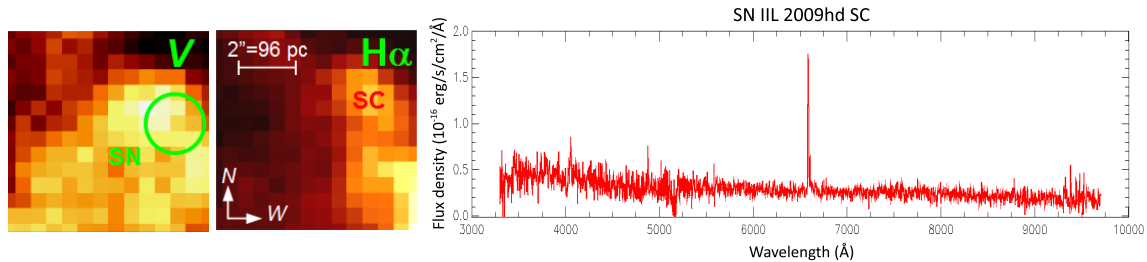}
\caption{IFU FoV and extracted host cluster spectra for SN 2009hd. Figure annotations are the same as in Figure \ref{sp70G}.}
\label{sp09hd}
\end{figure*}

With SNIFS we found that the explosion site appears to be clumpy with the presence of at least three distinct stellar populations. We extracted the spectrum of the H\texttt{II} region nearest to SN 2009hd, 1 arcsec west of the SN. Its continuum is rising towards blue wavelength with strong H$\alpha$ emission (Figure \ref{sp09hd}). Using N2 calibration we determined the metallicity of this cluster to be solar, and H$\alpha$ equivalent width yields age of 6.4 Myr. This corresponds to the lifetime of a 29.3 M$_\odot$ star.

\medskip
\begin{flushleft}
\textit{3. \objectname{SN 2009kr} site} \\
\end{flushleft}

\citet{eliasrosa10} and \citet{fraser10} independently reported the results of their studies on SN 2009kr. Using pre-explosion HST images, both studies managed to pinpoint the progenitor of the SN (positional uncertainty in the order of 0.02") and agree on the yellow supergiant nature of the star. However, \citet{fraser10} concluded that the initial mass of the progenitor is about 15 M$_\odot$, while \citet{eliasrosa10} determined the progenitor initial mass to be higher, around 18--24 M$_\odot$. This difference is caused by the different interpretations of the progenitor location on H-R Diagram. \citet{eliasrosa10} reported solar environment metallicity of 12+log(O/H) = 8.67 in PP04's N2 scale, but without clearly mentioning which part of the progenitor's environment.

\begin{figure*}
\plotone{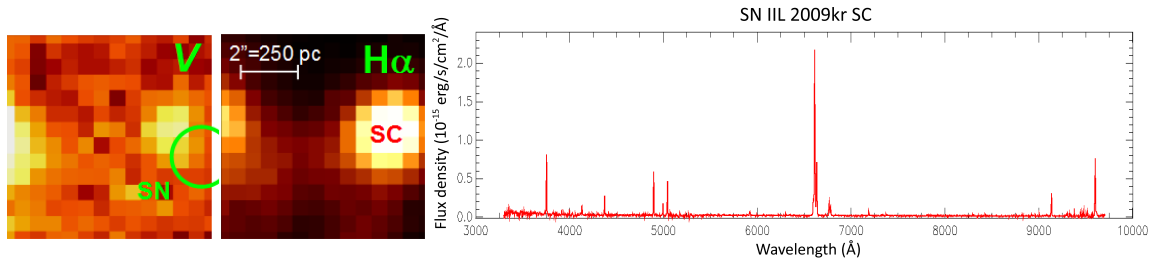}
\caption{IFU FoV and extracted host cluster spectra for SN 2009kr. Figure annotations are the same as in Figure \ref{sp70G}.}
\label{sp09kr}
\end{figure*}

The SN exploded in the outskirts of a large star cluster in the spiral galaxy NGC 1832. The SNIFS spectrum of the cluster (Figure \ref{sp09kr}) at a glance immediately shows the youthful nature of the cluster: faint continuum dominated by strong emission lines. We determined the metallicity of the cluster to be slightly lower than solar, 0.89 Z$_\odot$, using both O3N2 and N2 indices. The equivalent width of H$\alpha$ line indicates very young age of 3.3 Myr. This corresponds to the lifetime of a very massive star with mass in the order of $\sim117$~M$_\odot$. 

\medskip
\subsection{SNe II-P sites}

\medskip
\begin{flushleft}
\textit{1. \objectname{SN 1961I} site} \\
\end{flushleft}
SN 1961I appears to be not well studied. We could not find any reference reporting the characteristics nor the exact subtype of this SN except that it is a type II SN. We provisionally take this SN along with SN 1994L as type II-P, simply based on the observed fact that SNe II-P are the most frequent event within the type II SN classification \citep{smartt09}. The SN position measured on Palomar survey plates was shown to be accurate to within 2" by \citet{porter93}, as with SN Ic 1964L (c.f. Paper I).

\begin{figure*}
\plotone{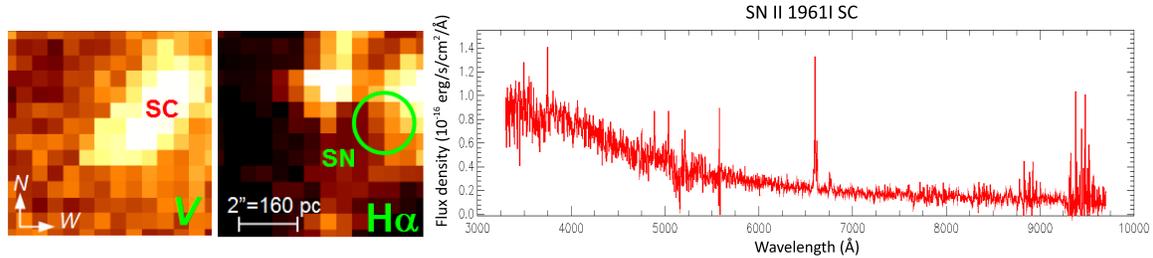}
\caption{IFU FoV and extracted host cluster spectra for SN 1961I. Figure annotations are the same as in Figure \ref{sp70G}.}
\label{sp61I}
\end{figure*}

\citet{boffi99} suspected that the bright patch at the explosion site may be an H\texttt{II} region or young open cluster. We confirmed this with SNIFS, finding that the object has a very blue continuum with Balmer emission lines (Figure \ref{sp61I}). We found out that the SN host cluster is 6.4 Myr old with 0.91 solar metallicity. Translated to stellar lifetime, at solar metallicity the age corresponds to a star with initial mass of 29.1 M$_\odot$. \\

\medskip
\begin{flushleft}
\textit{2. \objectname{SN 1994L} site} \\
\end{flushleft}
This object is not very well studied and only known for its type II classification. No further study has been found to secure a subtype for the classification of the object. As with SN 1961I, this SN was provisionally taken as a type II-P. The position of SN 1994L is probably accurate to 1", as estimated by \citet{vandyk92} for SNe of that era.

\begin{figure*}
\plotone{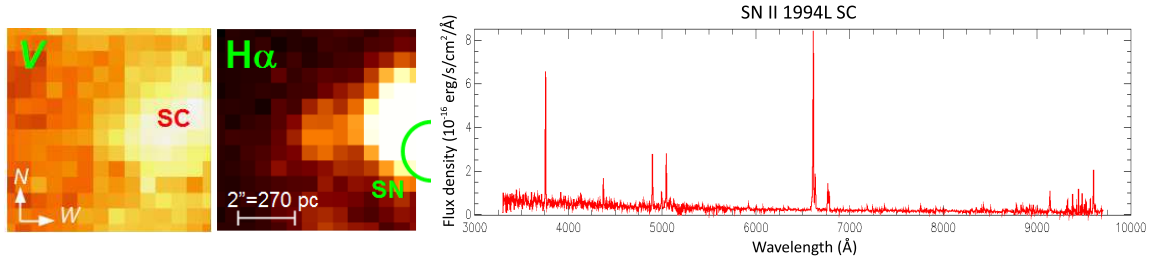}
\caption{IFU FoV and extracted host cluster spectra for SN 1994L. Figure annotations are the same as in Figure \ref{sp70G}.}
\label{sp94L}
\end{figure*}

The host galaxy NGC 2848 is an Sc spiral. No other SNe have been reported to explode in this galaxy. SN 1994L exploded in a bright cluster in the southern part of the galaxy. Our SNIFS pointing missed the explosion spot but a large portion of the cluster is within the field thus possible to be extracted (Figure \ref{sp94L}). Using SNIFS data we derived the host cluster age of nearly 5.0 Myr at 0.69 solar metallicity, which corresponds to the lifetime of a 45.9 M$_\odot$ star. 

\begin{flushleft}
\textit{3. \objectname{SN 1999gi} site} \\
\end{flushleft}
The progenitor of this SN has been searched for in HST pre-explosion images (positional uncertainty in the order of 0.02") but was not detected and only upper limits could be derived \citep[]{smartt09}. The upper mass limit for the progenitor star was derived to be 14 M$_\odot$.

\begin{figure*}
\plotone{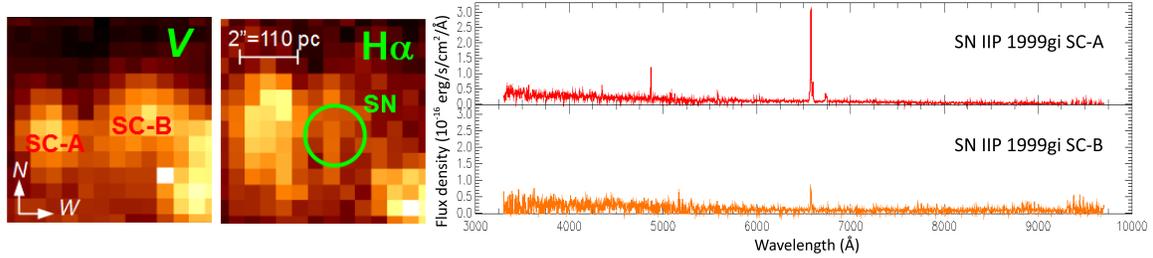}
\caption{IFU FoV and extracted host cluster spectra for SN 1999gi. Figure annotations are the same as in Figure \ref{sp70G}.}
\label{sp99gi}
\end{figure*}

Using SNIFS we observed the parent stellar population of SN 1999gi and the data shows that there are at least three clusters present at the explosion site within the SNIFS IFU field of view (Figure \ref{sp99gi}). For the cluster at the SN position (SC-B) we derived an age of 6.3 Myr, which corresponds to a progenitor mass of 29.4 M$_\odot$. We also measured a brighter cluster west of the SN host cluster and derived age of 5.5 Myr (corresponding to turn-off mass of 36.7 M$_\odot$). Both clusters have 0.79 and 0.72 solar metallicity, respectively. We could not measure the third cluster since it lies at the edge of the field of view thus only a small part of it is visible. \\

\medskip
\begin{flushleft}
\textit{4. \objectname{SN 1999gn} site} \\
\end{flushleft}
SN 1999gn is not well studied. It is a type II-P SN, and was suspected as a low-luminosity event \citep{pastorello04}. \citet{vandyk00} reported that the SN was observed during the Two Micron All Sky Survey (2MASS) operations and included in the 2MASS All-Sky Catalog of Point Sources \citep{cutri03}. The 2MASS coordinates of SN 1999gn agree with the coordinates from \citet{dimai99} within about 0.1" -- this value is assigned as the positional uncertainty of this SN.

\begin{figure*}
\plotone{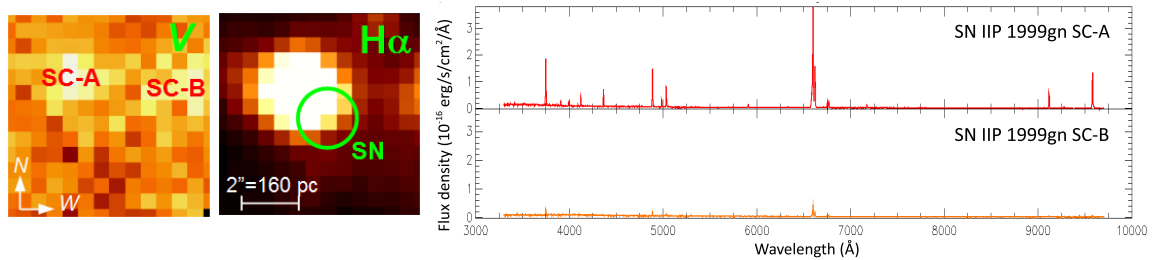}
\caption{IFU FoV and extracted host cluster spectra for SN 1999gn. Figure annotations are the same as in Figure \ref{sp70G}.}
\label{sp99gn}
\end{figure*}

SNIFS data shows that there are two clusters in the SN environment (Figure \ref{sp99gn}). The brighter one, the SN host cluster (SC-A), is very young with age of 3.3 Myr and metallicity of 1.07 solar. This age would imply a very massive progenitor star with initial mass in the order of $\sim117$~M$_\odot$. A neighbouring cluster was found to be older, 6 Myr with 1.02 solar metallicity. This age corresponds to the lifetime of 30.9 M$_\odot$ star.
\\

\medskip
\begin{flushleft}
\textit{5. \objectname{SN 2002hh} site} \\
\end{flushleft}
This interesting SN in NGC 6946 is well-studied. \citet{pozzo06} presented a $\sim1$ year photometric and spectroscopic monitoring of the SN done in optical and infrared. They inferred progenitor mass of 16--18 M$_\odot$ from the luminosity of [O\texttt{I}]$\lambda\lambda$6300,6364 line in the SN spectra. The self-obscuring progenitor of SN 2002hh has been suspected to be a massive M supergiant or an LBV that have undergone massive mass loss, inferred from the massive ($\gtrsim10$~M$_\odot$) dust+gas shell surrounding the SN \citep{barlow05}. This shell might have been produced by episodic ejection prior to SN explosion. However, \citet{meikle06} showed that the amount of circumstellar material present around the SN is likely to be smaller, $\sim3.6$~M$_\odot$. \citet{smartt09} derived progenitor upper mass limit of 18 M$_\odot$ from non-detection in pre-explosion images. Their positional uncertainty is in the order of 0.1"

With SNIFS we observe the explosion site of SN 2002hh and managed to capture the H\texttt{II} region north-east of the SN position (also present in \citet{meikle06}, Fig.~2). The spectrum of the object is dominated by strong emission lines indicative of a young stellar population. We  derived the metallicity of the host H\texttt{II} region as 1.05 solar using O3N2 and N2 indices, and an age of 5.8 Myr. The age corresponds to the lifetime of a 33.2 M$_\odot$ star.
\\
\begin{figure*}
\plotone{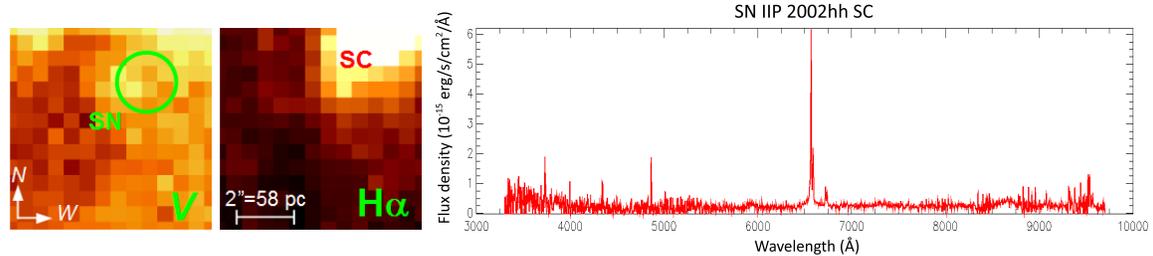}
\caption{IFU FoV and extracted host cluster spectra for SN 2002hh. Figure annotations are the same as in Figure \ref{sp70G}.}
\label{sp02hh}
\end{figure*}

\begin{figure*}
\plotone{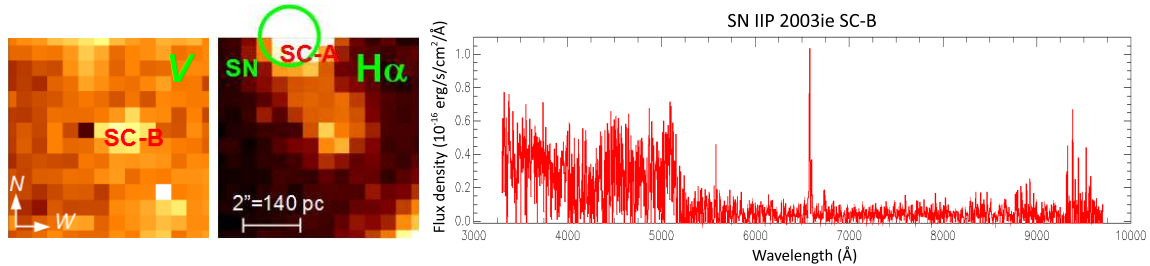}
\caption{IFU FoV and extracted host cluster spectra for SN 2003ie. Figure annotations are the same as in Figure \ref{sp70G}.}
\label{sp03ie}
\end{figure*}

\medskip
\begin{flushleft}
\textit{6. \objectname{SN 2003ie} site} \\
\end{flushleft}
There is lack of published study on this particular SN. \citet{smartt09} suggested that this SN may not be a normal type II-P, and derived progenitor upper mass limit of 24 M$_\odot$ from non-detection in pre-explosion images (positional uncertainty of 0.17"). The position of SN 2003ie coincides with a bright cluster with extended H$\alpha$ emission, but unfortunately our SNIFS pointing was inaccurate and only a small part of the host cluster fall inside the IFU FoV. We failed to extract the spectrum of this cluster (SC-A; see Figure \ref{sp03ie}). However, another faint cluster is visible at the south-east side of SC-A. This cluster appear to be associated with the extension of SC-A along the south-east direction. We managed to obtain the spectrum of this SC-B cluster and derived metallicity of 0.93 solar metallicity with age of 5.9 Myr. This corresponds with progenitor initial mass of 32 M$_\odot$. From our IFU reconstructed images we notice that SC-A is stronger than SC-B in H$\alpha$, so probably it is younger. This implies that the mass of the progenitor star may be higher than 32 M$_\odot.$
\\

\medskip
\begin{flushleft}
\textit{7. \objectname{SN 2004am} site} \\
\end{flushleft}
SN 2004am is not very well studied, but the position is coincident (positional uncertainty in the order of 0.1") with super star cluster L in M82 \citep{smartt09}, providing an interesting opportunity to derive its progenitor properties. A study by \citet{smith06} concludes the age of the cluster as $65^{+70}_{-35}$ Myr with limited SED fit from optical broadband photometry, while later work by \citet{lancon08} improved this estimate to $18^{+17}_{-8}$ Myr via near-infrared broadband photometry SED fitting. It is apparent that the red SED of the cluster could lead to misinterpretation of the cluster as an old stellar population. 

\begin{figure*}
\plotone{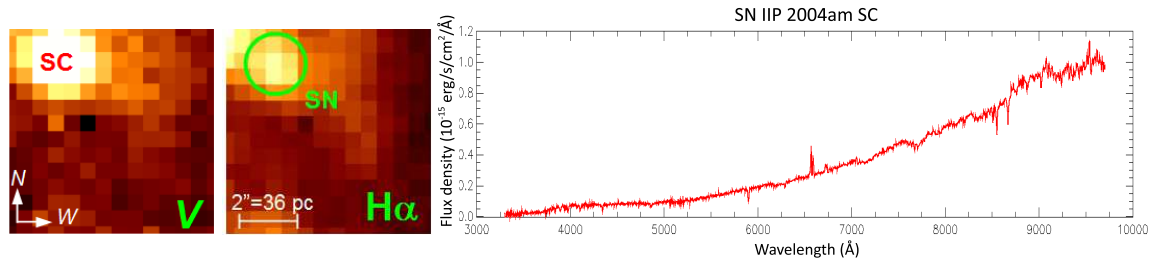}
\caption{IFU FoV and extracted host cluster spectra for SN 2004am. Figure annotations are the same as in Figure \ref{sp70G}.}
\label{sp04am}
\end{figure*}

Our SNIFS spectrum shows unambiguously that M82-L is a young cluster exhibiting prominent Balmer and Ca\texttt{II} triplet absorption lines, but severely reddened hence the red appearance of the SED (Figure \ref{sp04am}). H$\alpha$ shows up as an emission in the spectrum. We derived the N2 metallicity of the cluster as 1.35 solar. From H$\alpha$ emission equivalent width we determined age of the cluster as 12.6 Myr, while CaT EW yields age between 9 and 15 Myr with mean value of 12.8 Myr. These two measurements agree very well with each other and consistent with the $18^{+17}_{-8}$ Myr estimate of \citet{lancon08}. The mean H$\alpha$+CaT age of 12.7 Myr corresponds to the lifetime of a 15.8 M$_\odot$ star. 
\\

\medskip
\begin{flushleft}
\textit{8. \objectname{SN 2004dj} site} \\
\end{flushleft}
SN 2004dj exploded in the nearby spiral galaxy NGC 2403. The position coincides (positional uncertainty in the order of 0.1") with the star cluster Sandage 96 in one of the spiral arms of the host galaxy. Several authors have attempted to characterize the progenitor star based on the study of Sandage 96. \citet{maiz04} presented their study of Sandage 96 using published photometry and fitted the SED with Starburst99 SSP models to derive an age of 13.6 Myr, based on which they propose the 15 M$_\odot$ initial mass of the SN progenitor. Solar metallicity of the cluster is assumed in this work. Another estimate by \citet{wang05} yields cluster age of $\sim20$ Myr ($\sim12$~M$_\odot$ SN progenitor) at 0.4 solar metallicity. Further, \citet{vinko09} studied Sandage 96 in detail after SN 2004dj has faded and found that the age distribution of the stellar population within the cluster is bimodal, $\sim10-16$ Myr and $\sim32-100$ Myr. While it is likely that the older population are captured field stars, the younger population gives mass estimate for SN 2004dj progenitor as $\sim12-20$~M$_\odot$. The 10 Myr lower limit is supported by the lack of H$\alpha$ emission associated with Sandage 96, as revealed by narrow band H$\alpha$ imaging.

\begin{figure*}
\plotone{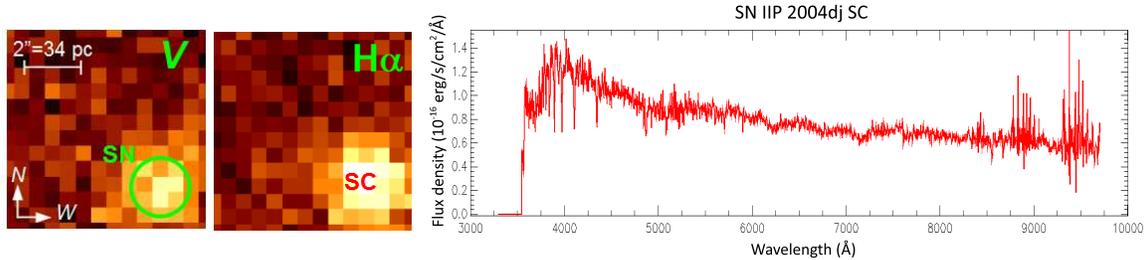}
\caption{IFU FoV and extracted host cluster spectra for SN 2004dj. Figure annotations are the same as in Figure \ref{sp70G}. The blue end of the spectrum is truncated due to DAR.}
\label{sp04dj}
\end{figure*}

Our SNIFS observation also shows that the cluster is dominated by continuum light without any detectable nebular emission throughout the instrument spectral response. The extracted spectrum confirms that the light of the cluster is mainly continuum emission produced by a young population (Figure \ref{sp04dj}). The overall spectrum is blue in color, showing prominent Balmer absorptions. No emission line is detected in the cluster spectrum. To determine metallicity, we used our observation of a nearby H\texttt{II} region, 35 arcsec eastward of Sandage 96 (corresponding to 595 pc at NGC 2403's distance). The metallicity of the H\texttt{II} region is subsolar, 0.33 Z$_\odot$, and adopted as the metallicity of Sandage 96. We used near-infrared Ca\texttt{II} triplet equivalent width as the age indicator. Using Starburst99 SSP model at 0.4 Z$_\odot$ we determined the age of Sandage 96 as 15.6 Myr, which corresponds to the lifetime of a 14.7 M$_\odot$ star at that corresponding metallicity. This result is consistent with previous studies which provided age estimates between $\sim10$ and 20 Myr.


\medskip
\begin{flushleft}
\textit{9. \objectname{SN 2005ay} site} \\
\end{flushleft}
SN 2005ay was discovered to be a subluminous SN II-P by \citet{tsvetkov06}. Based on light curve and spectral analysis, they suggested that the SN was produced by a progenitor with mass around the low end of SN II-P progenitor mass distribution.

\begin{figure*}
\plotone{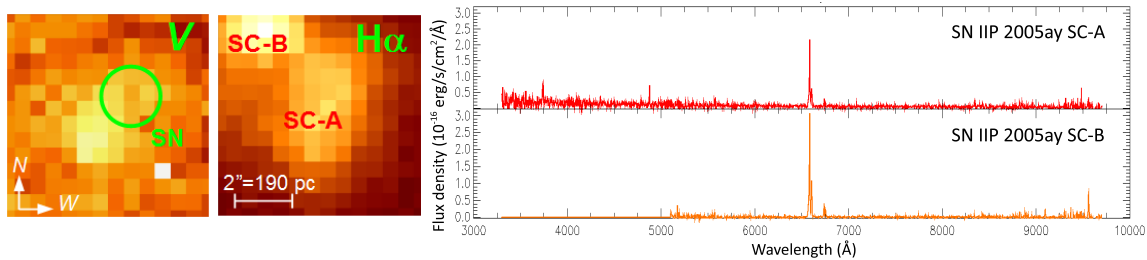}
\caption{IFU FoV and extracted host cluster spectra for SN 2005ay. Figure annotations are the same as in Figure \ref{sp70G}. The blue end of SC-B spectrum is truncated due to DAR.}
\label{sp05ay}
\end{figure*}

Our IFU data (Figure \ref{sp05ay}) shows that SN 2005ay was situated at the north-eastern edge of a fuzzy cluster (SC-A). The cluster was found to be of 0.81 solar metallicity from N2 determination, with age of 5.6 Myr. In the site there is a neighbouring cluster visible in H$\alpha$ only (SC-B). This turned out to be a very young H\texttt{II} region with age of 2.7 Myr at nearly solar metallicity, 0.95 Z$_\odot$. The age of SC-A corresponds to a high progenitor mass of 35.9 M$_\odot$.

\medskip
\begin{flushleft}
\textit{10. \objectname{SN 2008bk} site} \\
\end{flushleft}
SN 2008bk exploded in the spiral galaxy NGC 7793, and is the only SN ever recorded to explode in that galaxy. \citet{mattila08} reported their discovery of the progenitor star in high-quality pre-explosion VLT images. Their positional uncertainty is around 0.05". From optical and near-infrared photometry they derived the luminosity and colors of the star, which lead into a mass estimate of $\sim8.5$~M$_\odot$. Two years later, the disappearance of the purported progenitor star was confirmed and reported in \citet{mattila10}. This makes SN 2008bk one of the prime examples of SN with detected progenitor star in pre-explosion, well-characterized and confirmed disappearance after the explosion. \citet{vandyk12} also reported their analysis on their pre-explosion detection of the progenitor star, which was consistently determined to be between 8 and 8.5 M$_\odot$.

\begin{figure*}
\plotone{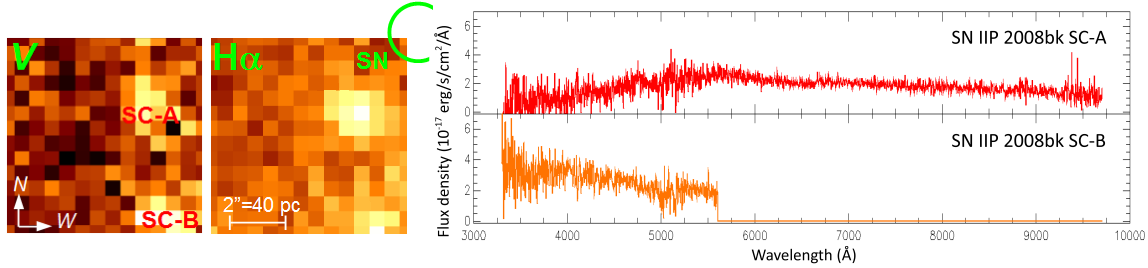}
\caption{IFU FoV and extracted host cluster spectra for SN 2008bk. Figure annotations are the same as in Figure \ref{sp70G}. The red part of SC-B spectrum was not obtained due to DAR.}
\label{sp08bk}
\end{figure*}

Our SNIFS observation of the explosion site unfortunately did not manage to cover the exact location of the SN due to the inaccuracy in pointing (Figure \ref{sp08bk}). However, two sources in the direction south-east of SN position were within the IFU field of view, thus may provide additional information about the immediate environment of the progenitor star. The northern object (SC-A) shows a relatively red continuum without any noticeable emission line. We only managed to obtain the blue part of the spectrum of SC-B, due to its position at the field edge. Nevertheless, it shows a markedly different appearance with the spectrum of SC-A. The continuum is rising towards bluer wavelengths, thus probably indicating a younger age. 

The absence of H$\alpha$ emission in the site suggests that the age of the environment is not very young, i.e. older than $\sim10-15$ Myr. This is consistent with the low-mass determination of SN 2008bk progenitor, whose $\sim8$~M$_\odot$ initial mass should correspond to the stellar lifetime in the order of $\sim30-40$ Myr. 

\section{Cluster membership probability and contamination estimation}
\label{contam}
To check the reliability of our result, we estimate the membership probability of the SN progenitor star to its host stellar population based on its projected position relative to the host cluster light profile. We also address the possibility of contamination from other clusters in the field. These clusters are supposedly fainter, hence invisible within our detection limit, and may have been the real parent cluster of the SN progenitor instead of the visible brightest clusters in the environment.

The light profile of the host cluster was determined by using the IRAF task \textit{imexamine}, and the FWHM was normalized by the seeing size at the time of observation. The separation, i.e. radial distance of the SN position from the cluster center, was then compared with the radial light profile of the cluster, to see where the SN position falls within the cluster light profile. The radial membership probability curve of each cluster was derived as the number of stars in the cluster at each radius; this is obtained by multiplying the light profile with the area of rings at different radii. The number density of star is assumed to be proportional to the light from the cluster. If the SN progenitors are located randomly within the host cluster, they should show concentration toward the peak probability. On the other hand, if the SN progenitors originated from the field thus have no association with the cluster, the separation distribution would not prefer any particular position.

\begin{figure}[Ht!]
\epsscale{1.}
\plotone{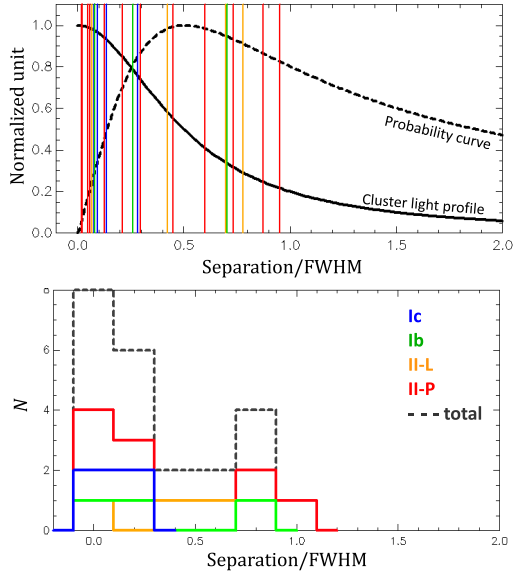}
\caption{All SNIFS SN-host cluster separations compared to normalized cluster light profile and the corresponding membership probability curve (upper panel); histogram of SN-host cluster separation (lower panel). SN positions are represented by vertical lines in the upper diagram. Color code indicates different SN types.}
\label{separ}
\end{figure}

In the upper panel of Figure \ref{separ} the separations of each SN in our SNIFS sample (from Paper I and Paper II) are plotted against the normalized host cluster light profile and the probability curve; the histogram is plotted in the lower panel. It is apparent that all SNe fall within 1$\times$FWHM radius from the cluster center, and the majority show concentration toward cluster center. On the other hand, the distribution of SN-cluster separation clearly does not follow the peak of the analytical probability curve around radius 0.5$\times$FWHM. This shows that the SN progenitors are concentrated towards the cluster center and does not randomly appear anywhere in the cluster, further suggesting physical association between SN progenitors and their respective host cluster. The histogram for each different types of SN show two distribution peaks, at cluster center and around 0.8$\times$FWHM, except for SNe Ic which are concentrated toward cluster center (Figure \ref{separ} upper panel). We note that positional uncertainty of the SN coordinates may affect this measurement. While for the most recent SNe, especially those whose progenitors were searched via direct imaging, the positions could be obtained with sub-arcsecond accuracy, for the older ones ($\sim$1990s) the uncertainty is in the order of 1" and may even reach around 10" for SNe observed before the $\sim$1980s \citep{vandyk92}. 

To address the possibility of contamination by fainter clusters, we performed a Monte Carlo simulation of SN-cluster separation by generating SNIFS FoV and adding clusters based on the cluster luminosity function of nearby galaxies. In nearby galaxies it was found that cluster luminosity function typically has slope with $\alpha \sim -2$ \citep{larsen02,mora09}, with number density of about 5--20 clusters/kpc$^2$. We applied this luminosity function to add invisible clusters into our fields, scaled by the number of actually detected clusters and assuming a somewhat overestimated number density of 20 clusters/kpc$^2$. Typically there is 1 dominant cluster accompanied by 2 fainter clusters in the simulated field of $360\times360$ pc, the median size of our observed fields. This represents the real observed situation where we usually detect 1 or 2 clusters in the SNIFS FoV, implying that there might be 2--4 more invisible clusters there according to the assumed cluster luminosity function. One can quickly compare the likelihood in harbouring SN progenitor star between the dominant cluster and the fainter ones, by assuming a cluster luminosity function for a population of identical star clusters in terms of IMF, age, metallicity and other physical properties so that the only difference is the number of stars within the cluster. With a cluster $\sim$2--3 magnitudes brighter than the detection limit -- the case of SNIFS sample -- the fainter clusters should have about 6--15 times less star since cluster luminosity is proportional to the number of stars. Even though the luminosity function dictates that there should be 2--4 of these faint clusters for each one bright cluster, in conclusion their combined likelihood to harbour a SN progenitor is still $\sim$30\% less than the bright cluster. Actually this would not even change the SN progenitor mass determination since same cluster age was assumed. If the fainter clusters are older (which is quite likely, since clusters tend to disperse thus becoming less luminous as they age \citep{fall05}), their likelihood of harbouring SN progenitor would become even smaller. Therefore, statistically it is more likely that the bright clusters are the real hosts of the SN progenitors. \citet{bastian06} estimated that cluster stars would still be physically associated to the cluster for 10--40 Myr. 

Even if the parent cluster is dispersing, the progenitor star would still not be too far away from the host cluster, considering the relatively short lifetime. With typical velocity dispersion within a cluster of few km/s ($\approx$ few pc/Myr), the progenitor star would still around within a few tens of pc away from the host cluster if it was not a runaway star. 
It is true that the possible distance traversed increases with longer lifetime (lower progenitor mass), but we can expect those to be still around 100-200 pc at most. Furthermore, if the progenitor movement is mostly radial or has significant radial component then it would appear even closer to the host cluster.

In the Monte Carlo simulation we generated two populations of SN--cluster association. The first population is fully random: the positions of clusters and SN are randomized within the simulated field thus no association whatsoever between the SN and the clusters. In the second population, the position of the SN is randomized around the 0.5$\times$FWHM of the brightest cluster's light profile (peak of the membership probability curve), thus representing the association between the two. We generated distribution models by combining these two populations with different proportions, and compare each one with the observed separation of SN--cluster. Kolmogorov-Smirnov (KS) tests were performed to check whether the model and the observed separation may belong to the same population.

\begin{figure}[Ht!]
\epsscale{1.}
\plotone{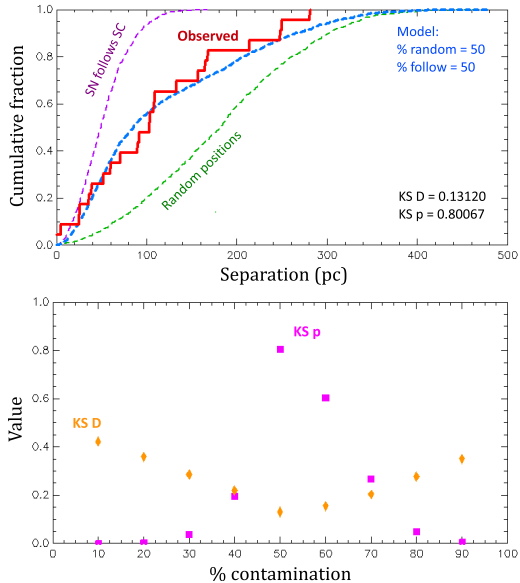}
\caption{Estimation of contamination by fainter clusters in the field. See text for explanation.}
\label{contamfig}
\end{figure}

Our simulation shows that the observed separation distribution is best represented by a model containing 50\% random SN-cluster population and 50\% associated SN-cluster population. In the upper panel of Figure \ref{contamfig} we plot the comparison between the observed distribution and the 50:50 model, which yields KS-test probability value of 80\% that both distributions may came from the same population. In the lower panel we show the KS probability values from different models containing random:associated populations, from 10:90 (association-dominated) to 90:10 (random-dominated), peaking at 50:50 composition. This shows that approximately 50\% of all SN-cluster pair might be real physical association, while the other 50\% may be just contamination from chance alignment. The consequence is discussed in the following section.

\section{Discussions}
\label{discu}

The importance of metallicity is believed to be paramount in the evolution of stripped-envelope (Ib/c) SNe. This is because a mechanism is thought to be necessary to remove the outer layers of the progenitor star to produce a hydrogen-deficient core-collapse SN. Metallicity-driven stellar wind is thought to be one viable scenario. In addition, removal of the envelope via close binary interaction is also thought as one viable scenario. However, in the case of SN type-II, there is still a significant amount of hydrogen envelope present at the time of explosion. Type II-L SN progenitors are presumed to have lost a large portion of their hydrogen envelope, but still retains some part of it before the explosion. 

Our measurement shows that on average, SN II-L progenitors has lower metallicity compared to II-P. The results are tabulated in Tables \ref{tabresl} and \ref{tabres}. We found the metallicity value of $0.81\pm0.22$ (RMS; standard error of the mean/SEM = $\sigma/\sqrt{N} = \pm0.12$) Z$_\odot$ for II-L progenitors and $0.88\pm0.28$ (SEM = $\pm0.09$) Z$_\odot$ for II-P. The difference is not significant, only within 0.4$\sigma$. As SNe II-L have lost more envelope compared to SNe II-P, this result is somehow contradictive to the expectation that SN II-L progenitors should have higher metallicity since higher metallicity will result in a more vigorous mass loss -- assuming a similar progenitor mass range. However, we note that this result is based on only a small number (three) of SN II-L progenitors.

We found that the average age of SN II-L host clusters are younger than II-P hosts, $4.3\pm1.8$ (SEM = $\pm1.0$) Myr compared to $7.6\pm4.2$ (SEM = $\pm1.5$) Myr. The difference is of $1.8\sigma$ significance. When the derived progenitor mass is compared, it is apparent that SNe II-L are produced by stars of higher mass compared to SNe II-P, as recent direct observation of progenitors suggest \citep[e.g.][]{eliasrosa10,eliasrosa11}. We found the mean initial mass for II-L progenitors is $84.1\pm47.8$ (SEM = $\pm15.9$)~M$_\odot$ while it is $39.3\pm30.8$ (SEM = $\pm3.4$)~M$_\odot$ for II-P ($1.5\sigma$ difference). The mean values are very high, even higher than the initial mass of SN Ib/c progenitors (see Paper I, but note the large sigma). If progenitors with mass greater than 100 M$_\odot$ are ignored, the mean value for SN II-P progenitor mass would decrease to $29.5\pm10.3$~M$_\odot$, similar with SN 2009hd progenitor mass of 29.3 M$_\odot$, the only SN II-L progenitor under 100 M$_\odot$. 

\begin{figure*}
\epsscale{1.1}
\plotone{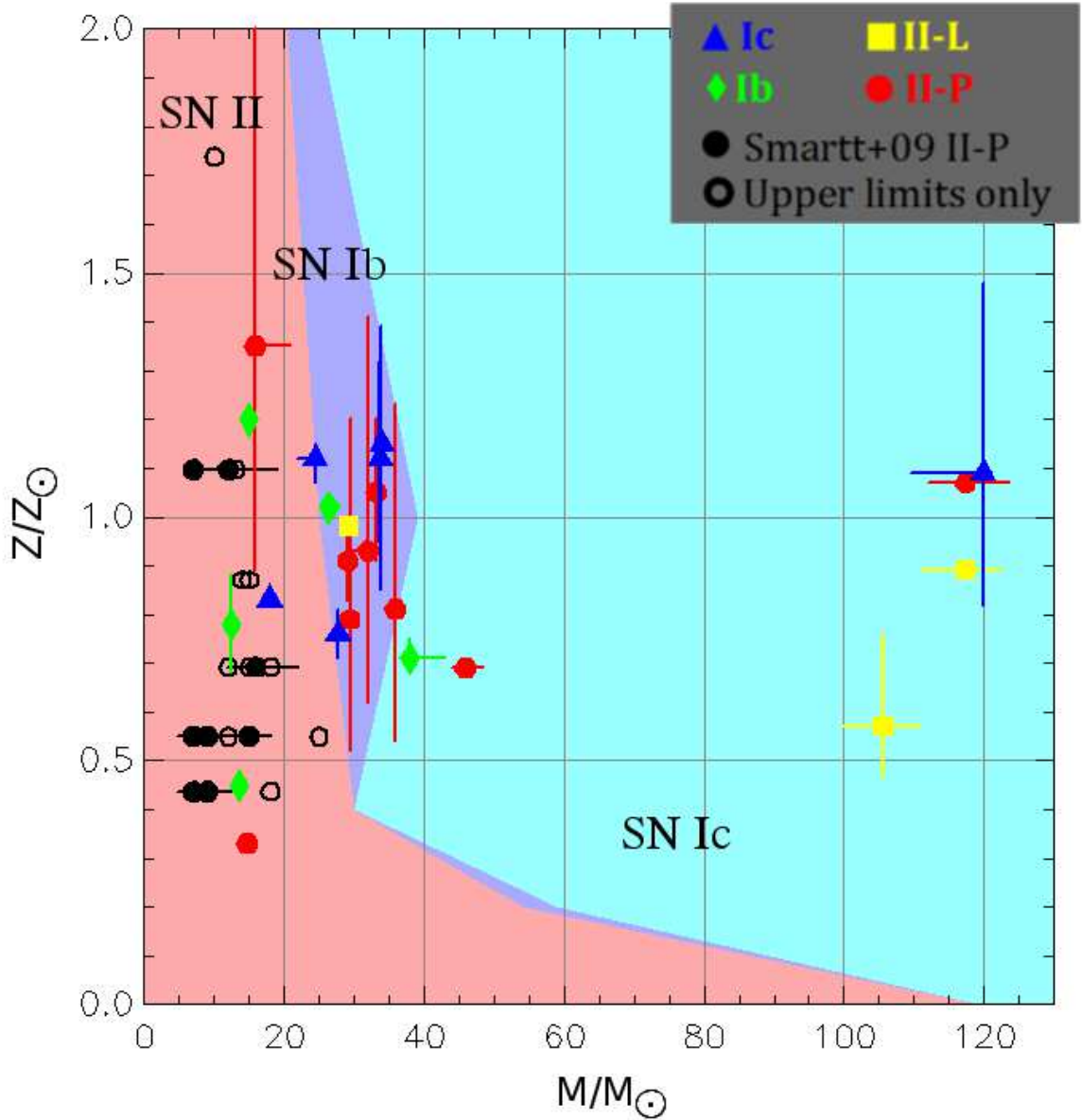}
\caption{SN II-P/II-L progenitors from SNIFS analysis overplotted on mass-metallicity space; II-P progenitors are indicated with filled red circles and II-L progenitors are indicated with yellow filled squares. Theoretical predictions of progenitors of each SN type by \citet{georgy09} are drawn as coloured regions. For comparison, SN II-P progenitors from \citet{smartt09} are plotted with black circles and hollow circles indicate progenitor mass upper limits only; note that for these data points metallicity was mostly determined using proxy, not direct measurement at the explosion site.}
\label{mzdiag}
\end{figure*}

In Figure~\ref{mzdiag} we plot the mass and metallicity determinations for SN progenitors in our sample, including the Ib/c ones presented in Paper I, on the mass-metallicity diagram of \citet{georgy09}'s model. It is apparent that the type-II progenitors are scattered all over the three regions on the diagram, even reaching the high-mass regions predicted for SN Ib/c progenitors. \citet{smith11frac} pointed out that some of the very massive stars may still retain their hydrogen envelopes by the time of SN explosion, resulting in type-II SNe.

The average progenitor mass value decreases drastically if we only consider the best cases only, i.e. SN within 150 pc of the parent cluster center (which indicates high association between cluster and SN) and with estimated progenitor mass not exceeding 100 M$_\odot$. Most of our clusters are smaller than this size (150 pc), and only a few are of approximately this size.

However, if we apply this criterion only SN 2009hd is left within the SNe II-L sample. With this criterion and excluding two cases where we could not recover the host cluster well (SNe 2003ie and 2008bk), the SN II-P progenitor mass would reduce into $24.4\pm8.6$ M$_\odot$, or $25.3\pm7.9$ M$_\odot$ if SN 2009hd is added to the SN II-P population. This average mass is rather similar to that of SN Ib ($22.6\pm12.1$~M$_\odot$) and  SN Ic progenitors ($27.5\pm6.7$~M$_\odot$) obtained in Paper I. If the combined populations of SNe Ib/c and SNe II-P/L are compared, the difference in initial mass is less than 0.1$\sigma$, signifying the similarity of the two populations.

In our contamination analysis, it was found that half of the sample of SN progenitor and host cluster may have been just a chance superposition. Considering the fact that all of our II-L progenitors (3/3) are massive ($\gtrsim25$~M$_\odot$) and more than half (7/10) are so for II-P progenitors, taking into account our estimate of 50\% contamination implies that at least some of SN II progenitors are massive stars comparable to SN Ib/c progenitors. If true, this conclusion poses a challenge to the current understanding of SN progenitors, where both theoretical and observational works are pointing toward sub-WR mass stars as SN II progenitors. However, already there is evidence that some massive stars may still retain their hydrogen envelope before exploding -- usually as a type IIn SN. \citet{smith11} derived an initial mass for type IIn SN 2010jl as larger than 30 M$_\odot$ from the observation of a luminous blue point source at the SN position before explosion -- either a star or the host star cluster, or combination of both, but all give similar constraints on the massive progenitor mass. \citet{kiewe12} suggested that typical SN IIn are produced by massive stars that have undergone luminous blue variable (LBV)-like mass loss events before the explosion. One compelling piece of evidence of an LBV progenitor of SN IIn is given by \citet{galyam09}; the explosion site of type IIn SN 2005gl was observed before and after the explosion, and it is evident that the blue source at the SN position disappeared after the explosion. This source was interpreted as the LBV progenitor of the SN, with initial mass of over $\sim50$~M$_\odot$. Our result may serve to provide more evidence to support the notion that some massive stars may still retain their hydrogen envelope prior to SN, producing a type II event. The light curve decline of type IIn SNe has been found to span a range from rapid decline similar to stripped SNe to plateau-like similar to SNe II-P \citep{kiewe12}. Recently, \citet{mauerhan13} suggested a new class of SN IIn-P, which show spectral signatures of a type IIn SN but with plateau-type light curve of SN type II-P. The progenitors of this class of SN may have been stars of around 8--10 M$_\odot$, or $\gtrsim25$ M$_\odot$ similar to some of our derived progenitor masses. Given the diversity of type IIn light curve, the lack of multi-epoch spectroscopic observation of type II SNe may lead into misidentification of the SN type, from type IIn into II-P or II-L based on the light curve only. We note that this may be the case for some SNe in our sample, especially the historic and poorly-observed ones. 

\begin{figure}[Ht!]
\epsscale{1.}
\plotone{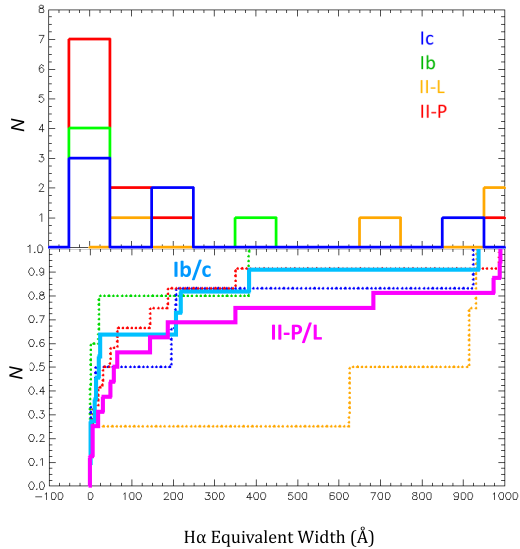}
\caption{Histogram of host cluster H$\alpha$ equivalent width (upper panel) and its cumulative distribution (lower), for each type of SN.}
\label{hahisto}
\end{figure}

Disregarding any physical interpretation, we plot the distribution of all host cluster H$\alpha$EW for each SN type in Figure \ref{hahisto}. It is apparent that the H$\alpha$EW distribution of SN II hosts is similar to the SN Ib/c ones. \citet{cedres02} show that the distributions of H$\alpha$EW of the H\texttt{II} regions in two nearby spiral galaxies, NGC 5457 and NGC 4395, peak around log(H$\alpha$EW) $\sim 3$. Considering this, it is possible that the objects with H$\alpha$EW $\gtrsim900\AA$ in Figure \ref{hahisto} are just outliers, and included in our sample as a result of random sampling. 

We also compare the result of our SN progenitor mass and environment age determination with previous studies using different methods, in Table \ref{tabcomp}. While our progenitor mass result is generally at odds with initial mass estimate from progenitor direct detection or nondetection, the derived environment age is generally consistent. This may indicate that SN progenitor mass determination via environment age may not result in the true progenitor mass. It is possible that the star formation history at the explosion site is not instantaneous and the SN progenitor emerged from older burst compared to younger burst that produced the dominant stellar population. We note that our method used in this work is sensitive to age about 20 Myr or younger (see Figure \ref{hacat}), thus insensitive to stellar populations harbouring stars less massive than $\sim12$~M$_\odot$. 

In Figure \ref{hacat} we also show the H$\alpha$EW evolution from Starburst99 for continuous star formation. If this continuous SFR was used to derive SSP age instead of the instantaneous one, the resulted age would lower significantly. This may push very massive progenitors to the lower mass region, but for the less massive ones this would bring them well into very old ages. In line with this, \citet{crowther13} pointed out that giant H\texttt{II} regions (with size of $\sim$100 pc or larger) may sustain several episodes of star formation, and the typical duty cycle of $\sim$20 Myr corresponds to the lifetime of a 12 M$_\odot$ star. We note that this star formation history caveat is present not only in our study but also in other similar works relying on stellar populations at SN explosion site \citep[e.g.][]{sanders12,leloudas11,levesque10}. We are currently trying to resolve this issue by increasing the sensitivity of the method toward older ($\gtrsim15$ Myr) stellar populations and mapping the star formation history of the explosion site, to be published elsewhere.


\begin{deluxetable*}{lcccccccrr}
\tabletypesize{\scriptsize}
\tablecaption{Results for SN II-L sites}
\tablewidth{0pt}
\tablehead{
\colhead{SN site} & \colhead{Object} & \colhead{Offset/"(pc)\tablenotemark{a}} & \colhead{(O3N2)} & \colhead{(N2)} & 12+log(O/H) & \colhead{Z/Z$_\odot$} & \colhead{H$\alpha$EW/\AA} & \colhead{Age/Myr} & \colhead{M$_{}$/M$_\odot$}
}
\startdata
1970G & SC* & 4.3 (150) & 8.54 & 8.31 & 8.43 & $0.57^{+0.19}_{-0.10}$ & $990.50\pm42.59$  & $3.41^{+0.05}_{-0.03}$  & $106_{+5}^{-6}$  \\
2009hd & SC* & 1.0 (48) & 8.65 & 8.64 & 8.65 & $0.98^{+0.00}_{-0.03}$  & $57.78\pm9.82$ & $6.37^{+0.08}_{-0.06}$ & $29.3_{+0.2}^{-0.1}$  \\
2009kr & SC* & 1.4 (170)  & 8.61 & 8.62 & 8.62 & $0.89^{+0.02}_{-0.00}$ & $973.80\pm83.75$& $3.26^{+0.05}_{-0.04}$  & $117_{+5}^{-6}$ \\
\enddata
\tablenotetext{a}{Offset between SN and approximate cluster center.}
\tablenotetext{*}{SN parent cluster}
\label{tabresl}
\end{deluxetable*}

\begin{deluxetable*}{lccccccccrr}
\tabletypesize{\scriptsize}
\tablecaption{Results for SN II-P sites}
\tablewidth{0pt}
\tablehead{
\colhead{SN site} & \colhead{Object} & \colhead{Offset/"(pc)\tablenotemark{a}} & \colhead{(O3N2)} & \colhead{(N2)} & 12+log(O/H) & \colhead{Z/Z$_\odot$} & \colhead{H$\alpha$EW/\AA} & \colhead{CaT EW/\AA} & \colhead{Age/Myr} & \colhead{M$_{}$/M$_\odot$}
}
\startdata
1961I & SC* & 0.9 (71) & 8.58 & 8.65 & 8.62 & $0.91^{+0.07}_{-0.08}$ & $49.31\pm8.38$ & -- & $6.43^{+0.11}_{-0.06}$ & $29.1_{+0.2}^{-0.4}$ \\
1994L & SC* & 1.8 (241) & 8.50 & 8.50 & 8.50 &  $0.69^{+0.00}_{-0.00}$ & $351.30\pm59.72$  & -- & $4.99^{+0.10}_{-0.14}$ & $45.9_{+2.6}^{-1.9}$ \\
1999gi & SC-A & 2.5 (143) & 8.52 & 8.52 & 8.52 & $0.72^{+0.00}_{-0.00}$ & $347.90\pm59.14$ & -- & $5.47^{+0.12}_{-0.23}$ & $36.7_{+2.3}^{-1.2}$ \\
1999gi & SC-B* & 1.6 (91) & -- & 8.56 & 8.56 & $0.79^{+0.41}_{-0.27}$ & $66.9\pm28.77$ & --&  $6.32^{+0.27}_{-0.17}$ & $29.4_{+0.5}^{-0.9}$ \\
1999gn & SC-A* & 1.4 (112) & 8.69 & 8.68 & 8.69 & $1.07^{+0.00}_{-0.02}$ & $988.3\pm88.95$ & -- & $3.26^{+0.04}_{-0.05}$ & $117_{+6}^{-5}$ \\
1999gn & SC-B & 1.6 (128) & 8.68 & 8.66 & 8.67 & $1.02^{+0.03}_{-0.02}$ & $116.70\pm12.84$  & -- & $6.03^{+0.11}_{-0.04}$ & $30.9_{+0.7}^{-0.8}$ \\
2002hh & SC* & 2.1 (122) & 8.74 & 8.62 & 8.68 & $1.05^{+0.15}_{-0.14}$ & $188.20\pm54.58$ & -- & $5.83^{+0.15}_{-0.14}$ & $33.2_{+1.4}^{-1.5}$ \\
2003ie & SC-A* & 0.7 (49) & -- & -- & -- & -- & -- & -- & $\lesssim6$ & $\gtrsim32$ \\
2003ie & SC-B & 5.5 (385) & -- & 8.63 & 8.63 & $0.93^{+0.48}_{-0.31}$ & $145.70\pm62.65$ & -- & $5.94^{+0.29}_{-0.16}$ & $32.0_{+1.7}^{-2.3}$ \\
2004am & SC* & 0.0 (0) & -- & 8.79 & 8.79 & $1.35^{+0.70}_{-0.46}$ & $6.09\pm0.55$ & $7.71\pm0.69$ & $12.70^{+1.81}_{-3.77}$ & $15.8_{+5.0}^{-1.4}$ \\
2004dj & SC* & 0.0 (0) &-- & -- & -- & -- & -- & $5.83\pm0.78$ & $15.60^{+0.10}_{-0.10}$ & $14.7_{+0.0}^{-0.1}$ \\
2004dj & H\texttt{II} reg & 35 (595) & 8.16 & 8.19 & 8.18 & $0.33^{+0.01}_{-0.01}$ & $1382.00\pm234.94$ & -- & $3.18^{+0.13}_{-0.07}$  & $\gtrsim120$ \\
2005ay & SC-A* & 2.2 (205) & -- & 8.57 & 8.57 & $0.81^{+0.42}_{-0.27}$ & $310.80\pm52.84$ & -- & $5.55^{+0.10}_{-0.12}$ & $35.9_{+1.2}^{-1.0}$ \\
2005ay & SC-B & 4.2 (391) & -- & 8.64 & 8.64 & $0.95^{+0.49}_{-0.32}$ & $1659.00\pm713.37$ & -- & $2.66^{+0.62}_{-1.15}$ & $\geq115$ \\
2008bk & SC-A & 6.0 (120) & -- & -- & -- & -- & -- & -- & $\gtrsim15$ & $\lesssim14$ \\
2008bk & SC-B & 9.0 (180) & -- & --  & --  & -- & -- & -- & $\gtrsim15$ & $\lesssim14$ \\
\enddata
\tablenotetext{a}{Offset between SN and approximate cluster center.}
\tablenotetext{*}{SN parent cluster}
\label{tabres}
\end{deluxetable*}

\begin{deluxetable*}{lcccccc}
\tabletypesize{\scriptsize}
\tablecaption{Mass and age of SN II progenitors from this work, compared with other determinations}
\tablewidth{0pt}
\tablehead{
\colhead{SN} & \colhead{Mass (this w.)}& \colhead{Age (this w.)} & \colhead{Mass (ref.)} & \colhead{Age (ref.)} & \colhead{Method (ref.)} & \colhead{Reference} 
}
\startdata
II-L 1970G & 106 M$_\odot$ & 3.41 Myr & -- & -- & -- & -- \\
II-L 2009hd & 29.3 M$_\odot$ & 6.37 Myr & $\lesssim20$~M$_\odot$ & -- & direct detection & \citet{eliasrosa11} \\
II-L 2009kr & 117 M$_\odot$ & 3.26 Myr & 18--24 M$_\odot$, & -- &  direct detection & \citet{eliasrosa10}, \\
 & & & 15 M$_\odot$ & & direct detection  & \citet{fraser10} \\
II 1961I & 29.1 M$_\odot$ & 6.43 Myr & -- & -- & -- & -- \\
II 1994L & 49.5 M$_\odot$ & 4.99 Myr & -- & -- & -- & -- \\
II-P 1999gi & 29.4 M$_\odot$ & -- & $\lesssim14$~M$_\odot$ & -- & nondetection & \citet{smartt09} \\
 & -- & 6.32 Myr & -- & 4 Myr & environment CMD & \citet{smartt09} \\
II-P 1999gn & $\sim120$~M$_\odot$ & 3.26 Myr & -- & -- & -- & -- \\
II-P 2002hh & 33.2 M$_\odot$ & 5.83 Myr & $\lesssim18$~M$_\odot$ & -- & nondetection & \citet{smartt09} \\
II-P 2003ie & $\sim32$ M$_\odot$ & $\sim6$ Myr & $\lesssim24$~M$_\odot$ & -- &  nondetection & \citet{smartt09} \\
II-P 2004am & 15.8 M$_\odot$ & 12.7 Myr & -- & $65^{+70}_{-35}$ Myr & SC SED & \citet{smith06} \\
 & --  & 12.7 Myr & -- &  $18^{+17}_{-8}$ Myr &  SC SED & \citet{lancon08} \\
II-P 2004dj & 14.7 M$_\odot$ & 15.6 Myr & 15 M$_\odot$ & 13.6 Myr & SC SED & \citet{maiz04} \\
 & 14.7 M$_\odot$ & 15.6 Myr & $\sim12$~M$_\odot$ & $\sim20$ Myr & SC SED & \citet{wang05} \\
& 14.7 M$_\odot$ & 15.6 Myr  & $\sim12$-20 M$_\odot$ & $\sim10$-16 Myr & SC SED \& CMD & \citet{vinko09} \\
II-P 2005ay & 35.9 M$_\odot$ & 5.55 Myr & $\sim9$~M$_\odot$ & -- & SN properties & \citet{tsvetkov06} \\
II-P 2008bk & $\lesssim14$~M$_\odot$ & $\gtrsim15$ Myr & $\sim8.5$~M$_\odot$, & -- & direct detection, & \citet{mattila08}, \\
 & & & 8--8.5 M$_\odot$ & & direct detection & \citet{vandyk12} \\
\enddata
\tablecomments{CMD: color-magnitude diagram; SC SED: star cluster spectral energy distribution}
\label{tabcomp}
\end{deluxetable*}

\section{Summary}
\label{summ}

In this study we investigate the mass and metallicity of type II-P and II-L SN progenitors by studying the parent stellar population. With integral field spectroscopy, the explosion sites of the SNe were observed spatially and spectrally. This enables us to obtain spectra of the stellar populations present at the site, and derive the metallicity via strong line method. The age of each stellar population was derived from H$\alpha$ and/or Ca triplet equivalent widths, compared to theoretical values from Starburst99 SSP model. We note that this kind of study is limited by the unknown star formation history of the site, and also in several cases the SN was not well documented and studied. 

We found that on average SN II-L explosion sites are younger  and less metal rich compared to II-P progenitors. We estimate that about 50\% of our SN--parent cluster samples are chance alignments, leaving the other 50\% part as real physical associations. This estimate implies that at least some of the SN II-P/II-L progenitors were massive stars comparable to single SN Ib/c progenitors ($\gtrsim25$~M$_\odot$). The distribution of H$\alpha$ equivalent width of SN II host clusters also show similarity with that of SN Ib/c hosts. While this result may not be compatible with some of the contemporary theoretical models and observational results which point that SN II-P and II-L progenitors are stars less than $\sim25$~M$_\odot$, it does not contradict the notion that a massive star above the Wolf-Rayet mass limit could explode while still retaining its hydrogen envelope, in agreement with \citet{smith11frac}.



\acknowledgments
We acknowledge the anonymous referee for helpful comments and suggestions.
H.K. acknowledges generous support from the Japanese government MEXT (Monbukagakusho) scholarship. Useful help from R. Pain, S. Rodney, and P. Weilbacher on working with datacubes is appreciated.
We thank G. Leloudas for carefully reading the draft and providing important comments. We also thank J. Sollerman and F. Taddia for helpful comments on the draft of the manuscript.
This work was supported in part by a JSPS core-to core program "International Research Network for Dark Energy" and by JSPS research grants. 
This work is based on the data by using the University of Hawaii 88-inch Telescope (UH88), for which the telescope time was afforded by the funding from National Astronomical Observatory of Japan.
The work of K.M. is supported by World Premier International Research Center Initiative (WPI Initiative), MEXT, Japan, and Gant-in-aid for Scientific Research (23740141).
G.A. was supported by the Director, Office of Science, Office of High Energy Physics, of the U.S. Department of Energy under Contract No. DE-AC02-05CH11231. SNIFS on the UH 2.2-m telescope is part of the Nearby Supernova Factory II project, a scientific collaboration among the Centre de Recherche Astronomique de Lyon, Institut de Physique Nucl\'eaire de Lyon, Laboratoire de Physique Nucl\'eaire et des Hautes Energies, Lawrence Berkeley National Laboratory, Yale University, University of Bonn, Max Planck Institute for Astrophysics, Tsinghua Center for Astrophysics, and the Centre de Physique des Particules de Marseille.
This research has made use of the SIMBAD database and ALADIN, operated at CDS, Strasbourg, France.
This research has made use of the NASA/IPAC Extragalactic Database (NED) which is operated by the Jet Propulsion Laboratory, California Institute of Technology, under contract with the National Aeronautics and Space Administration.
The authors wish to recognize and acknowledge the very significant cultural role and reverence that the summit of Mauna Kea has always had within the indigenous Hawaiian community.  We are most fortunate to have the opportunity to conduct observations from this mountain. 
\\



{\it Facilities:} \facility{UH88 (SNIFS, OPTIC)}.





\clearpage

\end{document}